\documentclass{ifacconf}

\usepackage{natbib}
\usepackage{amsmath,amssymb,amsfonts,mathrsfs} % Regroupé
\usepackage{graphicx}
\usepackage{mathtools}
\usepackage{xcolor}
\usepackage{enumerate}
\usepackage[percent]{overpic}
\usepackage{comment}
\usepackage{url}
\usepackage[dvipsnames]{xcolor}

\usepackage{mathrsfs} % Very curly L for loss

\allowdisplaybreaks
\usepackage{rotating}
\newcommand{\solutionswitch}[2]{#2} % For Arxiv
\newtheorem{proposition}{Proposition}
\newtheorem{assumption}{Assumption}

\newtheorem{remark}{Remark}

\usepackage{tabularx,ragged2e,booktabs} % caption
\def\RR{{\mathbb R}}    % field of real number
\def\X{{\mathcal{X}}}

\begin{document}

\begin{frontmatter}

\title{Learning a Contracting KKL-observer with Local Optimal Guarantees}
\thanks[footnoteinfo]{This work was partially supported by NCCR Automation, a National
Centre of Competence in Research, funded by the Swiss National
Science Foundation (grant number 51NF40\_225155), and by the ANR under grant OLYMPIA ANR-23-CE48-0006. Corresponding author: C.L. Galimberti (\texttt{clara.galimberti@supsi.ch}).}

\author[idsia]{Clara L. Galimberti}
\author[lyon1]{Johan Peralez}
\author[lyon1]{Daniele Astolfi}
\author[lyon1]{Vincent Andrieu}
\author[lyon1]{Madiha Nadri}

\address[idsia]{Dalle Molle Institute for Artificial Intelligence (IDSIA), SUPSI, Via la Santa 1, 6962 Lugano-Viganello, Switzerland}
\address[lyon1]{Universit\'e Claude Bernard Lyon 1, CNRS, LAGEPP UMR 5007, 43 boulevard du 11 novembre 1918, F-69100, Villeurbanne, France}

\begin{abstract}
The Kazantzis-Kravaris-Luenberger (KKL) observer provides a general framework for nonlinear state estimation by immersing the system dynamics into a stable linear or nonlinear latent dynamics. However, the performance of KKL observers relies heavily on the specific choice of these latent dynamics, which is often heuristic. This paper proposes a methodology to learn a KKL observer that combines global stability guarantees with local optimality. We derive a condition on the latent dynamics such that the observer locally mimics the behavior of a Minimum Energy Estimator (Mortensen observer). We then employ Deep Learning to approximate the KKL transformation and the latent dynamics, using neural network architectures that 
structurally enforce the contraction property. The proposed strategy is validated through numerical simulations on nonlinear benchmarks, demonstrating 
a good performance in the presence of state and measurement noise.
\end{abstract}


\end{frontmatter}

\section{Introduction}
\label{sec:intro}

Accurate state estimation is a cornerstone of modern control, monitoring, and fault detection systems where only partial measurements of the state are available. For nonlinear systems, the Kazantzis-Kravaris-Luenberger (KKL) theory provides a powerful framework by immersing the nonlinear dynamics into a stable latent space \citep{andrieu2006existence,brivadis2023further,pachy2024existence}. The design relies on two steps: defining a contractive latent dynamics driven by the measured output, and reconstructing the state by inverting the immersion map.

Constructing KKL observers requires solving a nonlinear partial differential equation (PDE), which is often analytically intractable. To overcome this hurdle, recent research has successfully employed machine learning---particularly neural networks---to approximate the solution of the KKL PDE from data or models \citep{ramos2020numerical, peralez2021deep, peralez2024deep, niazi2022learning, miao2023learning, buisson2023towards}.
These approaches have significantly broadened the practical applicability of KKL observers, enabling their deployment on systems where analytical solutions are out of reach.

Despite these computational advances, a fundamental open question remains: \textit{how to select the latent dynamics}. This choice is critical as it dictates the overall evolution of the estimation error, governing both the transient response and steady-state performance. 
The goal of this work is therefore to 
provide a principled characterization of these dynamics, with a specific focus on local behavior, which in turn determines the (steady-state) observer's sensitivity to measurement noise and unmodeled state perturbations. Building upon Mortensen's observer
\citep{Mortensen1968}—which defines the optimal, albeit non-implementable, estimator—we first derive an approximated observer dynamics. While this approximation remains practically non-implementable, it establishes a benchmark for optimal local behavior when the estimation error is small.
Consequently, we develop a learning strategy based on Deep Learning to construct a KKL observer that locally matches this optimal approximated dynamics. Finally, we demonstrate the effectiveness of our proposed methodology through numerical simulations.

\solutionswitch{Proofs and supplementary material are available at \cite{galimberti2026}.}{All the proofs can be found in the Appendix.}

In the following, we denote with $|\cdot|$ the standard Euclidean norm and with $\|\cdot\|_{P}^2 = x^\top P x$ the norm induced by a positive definite matrix $P\succ 0$.

\section{Preliminaries on  KKL observers}
\label{sec:problem}
We consider a nonlinear system subject to state disturbances and measurement noise:
\begin{subequations}
\label{eq:sys}
\begin{align}
    \dot{x}(t) &= f(x(t)) + w(t), \quad x(0)=x_0, \label{eq:sys_x}\\ 
    y(t) &= h(x(t)) + v(t), \label{eq:sys_y}
\end{align}
\end{subequations}
where $x(t)\in\RR^{n_x}$ is the state trajectory and $y(t)\in \RR^{n_y}$ is the measured output. The signals $w: \RR_{\ge 0} \to \RR^{n_x}$ and $v: \RR_{\ge 0} \to \RR^{n_y}$, both in $L^\infty_{\rm{loc}}(\RR_{\geq0})$,\footnote{A function belongs to $L^\infty_{\mathrm{loc}}(\mathbb{R}_{\geq 0})$ if it is essentially bounded on every compact subset of $\mathbb{R}_{\geq 0}$.} represent unknown state disturbances and measurement noise, respectively. The functions $f$ and $h$ are assumed sufficiently smooth.

We define a compact set $\X \subset \RR^{n_x}$ representing the domain of interest where the state estimation is performed. Due to the presence of unmodeled disturbances $w$, we do not assume $\X$ to be forward invariant, but our analysis focuses on time intervals where $x(t) \in \X$.

We employ the KKL framework by introducing a 
latent dynamical system:
\begin{equation}\label{eq:phi_dynamics}
    \dot z(t) = \varphi(z(t),y(t)) ,
    \quad  z(0)=z_0,
\end{equation}
where $z(t) \in \RR^{n_z}$, with $n_z \ge n_x$, is latent state trajectory. The vector field $\varphi:\RR^{n_z\times n_y}\to\RR^{n_z}$
is a $C^1$ function parameterized to ensure robust stability property as follows.

\begin{assumption}[Contraction KKL map]\itshape
\label{ass:contraction}
There exist a symmetric positive definite
matrix $P$ and a positive real number $\varepsilon>0$
satisfying for all $(z,y)\in \RR^{n_z}\times \RR^{n_y}$
\begin{equation}
\label{eq:contraction_latent_dynamics}
P \dfrac{\partial \varphi }{\partial z}(z,y)
+
\dfrac{\partial \varphi }{\partial z}(z,y)^\top P \preceq - \varepsilon I.
\end{equation}
Furthermore, 
$\varphi$ is Lipschitz continuous with respect to the input $y$, with constant $L_{\varphi, y}$.
\end{assumption}

This implies that, under Assumption~\ref{ass:contraction},
the $z$-dynamics is incrementally exponentially stable.
Specifically, for any initial conditions $z_a, z_b$ and any input signal $y(\cdot)$,
the corresponding  trajectories 
$z_a(t),z_b(t)$
to \eqref{eq:phi_dynamics}
satisfy
\begin{equation*} %\label{eq:deltaGES_phi_dynamics}
    |z_a(t) - z_b(t)| \le \kappa e^{-\lambda t} |z_a - z_b|, \quad \forall t \ge 0,
\end{equation*}
where $\lambda, \kappa > 0$ are constants depending on $P$ and $\varepsilon$.

Next, we introduce the intertwining PDE associated with the disturbance-free system.
A solution to this PDE, denoted by $T: \X \to \RR^{n_z}$, is referred to as the KKL-\textit{transformation}: % TODO \clara{(or immersion map)}:
\begin{equation}\label{eq:PDE}
    \frac{\partial T}{\partial x}(x)f(x) = \varphi(T(x),h(x)), \quad \forall x \in \X.
\end{equation}
The observer design relies on the following key assumption.

\begin{assumption}[Injectivity]\label{ass:injectivity}\itshape
    There exists a smooth map $T$ solving \eqref{eq:PDE}, 
    with bounded Jacobian on $\X$, namely 
    $\sup_{x \in \X} |\frac{\partial T}{\partial x}(x)| \le L_T$,
    which admits a sufficiently smooth left-inverse map $\tau: \RR^{n_z} \to \RR^{n_x}$ satisfying
\begin{equation}\label{eq:tau_T_x}
        \tau(T(x)) = x, \qquad \forall \, x\in \X.
    \end{equation}
    Furthermore, $\tau$ is globally Lipschitz continuous with constant $L_\tau$.
\end{assumption}

\begin{remark}
    The existence of the left-inverse map $\tau$
    is generically ensured under some observability conditions and constraints on the dimension of the latent space $n_z$. We refer the reader to \cite{pachy2024existence,brivadis2023further,andrieu2006existence}   for more precise details. We remark that a trivial case can also be obtained by selecting $n_x= n_z$ and $\varphi(z,y) =f(z) +k(y  -h(z))$, following standard observer design in the 
    \textit{original coordinates}, see, e.g., 
    \cite{bernard2022observer}. 
\end{remark}

Under the previous assumptions, 
the KKL-observer is defined as 
\begin{equation}\label{eq:KKL-observer}
    \dot z(t) = \varphi(z(t),y(t)), \quad 
    \hat x(t) = \tau(z(t)), \quad z(0)=z_0.
\end{equation}
It satisfies the following 
robust convergence property formalized in the next proposition.

\begin{proposition}[Robust Estimation]
\label{prop:iss} 
    Consider the system~\eqref{eq:sys} and the observer \eqref{eq:KKL-observer}.
    Under Assumption~\ref{ass:contraction} and Assumption~\ref{ass:injectivity}, for any trajectory $x(t)$ remaining in $\X$ and for any $(z_0,w,v)$ in $\RR^{n_z}\times L^\infty_{\rm{loc}}(\RR_{\geq0})\times L^\infty_{\rm{loc}}(\RR_{\geq0})$, the estimation error satisfies,
    for all $t\geq0$:
    \begin{multline}\label{eq_ISS}
        |\hat{x}(t) - x(t)| \le L_\tau \kappa e^{-\lambda t} |z_0 - T(x_0)| \\
        + \frac{L_\tau \kappa}{\lambda} \sup_{s \in [0, t]} \left( L_{\varphi, y}|v(s)| + L_T |w(s)| \right).
    \end{multline}
\end{proposition}

\begin{remark}
    Note that inequality \eqref{eq_ISS} is not per se an ISS bound since on the right hand side of the inequality it is not $\hat x(0)-x(0)$. This property is in fact an asymptotic gain property (see \cite{bernard2022observer}).
\end{remark}

\section{Local behaviour of an observer}

\subsection{Problem Formulation}
\label{sec:problem_formulation}

The KKL framework provides a powerful existential result: for a broad class of nonlinear systems, an observer exists in the form of a contractive latent system \eqref{eq:phi_dynamics} coupled with an immersion map $T$. However, the theory leaves the selection of the latent dynamics $\varphi$ largely unconstrained. In practice, $\varphi$ is often chosen heuristically (e.g., linear diagonal dynamics with arbitrary eigenvalues), which may lead to poor performances under measurement noise.

Observer design typically involves balancing two conflicting objectives:
\begin{enumerate}[(i)]
    \item \textbf{Transient behavior:} Ensuring fast global convergence from arbitrary initial conditions. In the KKL framework, this is guaranteed by the contraction rate of $\varphi$ (Assumption \ref{ass:contraction}).
    \item \textbf{Steady-state behavior:} Minimizing the sensitivity to measurement noise and disturbances once the estimation error is small.
\end{enumerate}

While KKL observers naturally address the first objective through global contraction, their local noise rejection properties depend non-trivially on the interplay between the chosen $\varphi$ and the resulting geometry of the immersion $T$.
The core idea of this work is to leverage the degrees of freedom in choosing $\varphi$ to impose a specific \textit{optimal} behavior in the steady-state regime, without compromising global stability.

To achieve this, we must first characterize how the specific choice of the pair $(\varphi, T)$ influences the estimation error dynamics locally around the invariant manifold. This analysis is the subject of the following subsection.

\subsection{First-order Approximation}
\label{sec:first_order_approx}

In this section, we derive a first-order approximation of the dynamics of the reconstructed state $\hat{x}$. Crucially, the following analysis relies solely on the algebraic coupling provided by the PDE \eqref{eq:PDE} and the inversion map $\tau$, as stated in Assumption~\ref{ass:injectivity}, without requiring the contraction property (Assumption~\ref{ass:contraction}).
This distinction is fundamental: it implies that the local behavior characterization derived here applies to any system immersed into a latent dynamics, even if global stability guarantees are missing. This justifies our subsequent analysis of the Mortensen observer, treating it as a specific instance of an immersion-based observer structure.

We focus on the evolution of $\hat{x} = \tau(z)$ for small perturbations around the manifold $\mathcal{M} = T(\mathcal{X})$, i.e., when the latent state $z$ is close to $T(x)$. We require the following regularity bounds on the maps.

\begin{assumption}[Regularity Bounds]
\label{ass:bounds}\itshape
There exist positive constants $L_{\varphi,1}, L_{\varphi,2}$ and $L_{\tau,1}, L_{\tau,2}, L_{\tau,3}$ such that for all $(z,y)$ in $\RR^{n_z}\times h(\X)$:
\begin{align*}
    \left| \frac{\partial \varphi}{\partial z}(z,y) \right| &\le L_{\varphi,1}, \quad
    \left| \frac{\partial^2 \varphi}{\partial z^2}(z,y) \right| \le L_{\varphi,2}, \\
    \left| \frac{\partial^{(k)} \tau}{\partial z^{(k)}}(z) \right| &\le L_{\tau,k}, \quad \text{for } k=1,2,3.
\end{align*}
\end{assumption}

\begin{proposition}[Structural Local Dynamics]\label{prop:local_dyn_e}%
~~Consider \\the system \eqref{eq:sys} with $w=0$ and $v=0$, and a triplet $(\varphi, T, \tau)$ satisfying Assumption~\ref{ass:injectivity} and Assumption~\ref{ass:bounds}. Let $e = z - T(x)$ be the latent estimation error.
For any $x \in \mathcal{X}$, the dynamics of the reconstructed state $\hat{x} = \tau(z)$ satisfy:
\begin{equation}\label{eq:dyn_obs_exact_e}
    \dot{\hat x} 
    = f(\hat x) + \Psi(\hat x) (x-\hat x) 
    + \mathcal{R}(x, e),
\end{equation}
where the gain matrix $\Psi(\hat{x}) \in \RR^{n_x \times n_x}$ is defined as 
$\Psi(\hat x) = - \psi(\hat x)\frac{\partial T}{\partial x}(\hat x)$,
with $\psi(\hat{x}) \in \RR^{n_x \times n_z}$ given by:
\begin{align*}
    \psi(\hat{x}) &= - \frac{\partial f}{\partial x}(\hat{x}) \frac{\partial \tau}{\partial z} (T(\hat x))
       + \frac{\partial \tau}{\partial z}(T(\hat x)) \frac{\partial \varphi}{\partial z} (T(\hat x), h(\hat x)) \\
       & \quad + \frac{\partial^2 \tau}{\partial z^2}(T(\hat x)) \varphi (T(\hat x) , h(\hat x)).
\end{align*}
The remainder term $\mathcal{R}(x, e)$ satisfies the bound:
\begin{equation}\label{eq_BoundRemainder}
     |\mathcal{R}(x, e)| \le c_1 |e|^2 + c_2 |z - T(\hat{x})| + c_3 |z - T(\hat{x})|^2,
\end{equation}
where $c_1$, $c_2$, and $c_3$ are positive constants depending on the bounds of Assumption \ref{ass:bounds}.
\end{proposition}

The bound \eqref{eq_BoundRemainder} highlights two distinct sources of approximation error. 
The term proportional to $|e|^2$ represents the standard linearization error. 
%The term proportional to $|z - T(\hat{x})|$ represents 
The terms involving $|z - T(\hat{x})|$ and $|z - T(\hat{x})|^2$ represent
a ``transversal'' error, measuring the distance of the latent state $z$ from the invariant manifold $\mathcal{M} = T(\mathcal{X})$.

While this structural decomposition holds for any immersion, its practical relevance relies on the stability of the latent dynamics.
In our design strategy (Section \ref{sec:learning_method}), we will explicitly enforce the contraction property (Assumption \ref{ass:contraction}). Consequently, in the absence of disturbances, $z(t)$ will converge exponentially to $T(x(t))$, causing the transversal error $|z - T(\hat{x})|$ to vanish asymptotically.
This justifies our learning objective: we aim to optimize the effective gain $\Psi(\hat{x})$ to ensure optimal local behavior \emph{on the manifold}, relying on the global contraction to naturally drive the system state towards the region where this approximation holds valid.

\section{Definition  of local optimal behavior}
\label{sec:optimal_behavior}

In this section, we define the target dynamics for the local behavior of our observer. Our goal is to design an observer that behaves, locally around the estimated trajectory, like a Minimum-Energy Estimator (MEE)
\citep{Mortensen1968},  without the computational burden of solving the Hamilton-Jacobi-Bellman (HJB) equation in real-time.

\subsection{The Minimum Energy Estimator}
\label{subsec:mortensen}
The MEE provides an optimal state estimate by selecting the trajectory that best explains the measured output $y$ while minimizing the energy of the unknown disturbance.
Given an initial condition $x_0$ and a disturbance $w \in \mathcal{L}^2([0, t], \mathbb{R}^{n_x})$, the associated cost is
\begin{equation}
\label{eq:cost_functional}
\begin{split}
    J(x_0, w, t | y) = \frac{1}{2} \int_0^t \bigg( & \|y(s)-h(X(x_0, s; w))\|_{\mathbf{R}^{-1}}^2 \\
    & + \|w(s)\|_{\mathbf{Q}^{-1}}^2 \bigg) ds,
\end{split}
\end{equation}
where $X(x_0, s; w)$ denotes the %solution of \eqref{eq:sys_x}.
unique solution to system \eqref{eq:sys_x}, assumed to exist for all $s \ge 0$, initialized in state $x_0$ at time $s=0$ and driven by the disturbance signal $w$.

The optimal estimate at time $t$ is the endpoint of the minimizing trajectory:
\begin{equation*}
    \hat{x}^*(t) = X(x_0^*, t; w^*),
    \quad 
    (x_0^*, w^*) = \underset{x_0, w}{\arg \min} \,\, J(x_0, w, t | y).
\end{equation*}
This variational problem admits a dynamic programming formulation.
Let $\mathcal{V}_y(\chi, t)$ denote the minimum energy required to reach the state $\chi$ at time $t$. 
The optimal estimate satisfies $\hat{x}^*(t) = \arg \min_z \mathcal{V}_y(\chi, t)$. Minimizing the associated Hamiltonian yields the relation between the optimal disturbance and the value function gradient:
\begin{equation*}
    w^*(t) = - \mathbf{Q} \nabla \mathcal{V}_y(\hat{x}^*(t), t)^\top,
\end{equation*}
where the value function $\mathcal{V}_y$ satisfies the HJB equation with respect to the spatial variable $\chi$:
\begin{equation}
    \label{eq:HJB}
    \frac{\partial \mathcal{V}_y}{\partial t} 
    =
    - \frac{\partial \mathcal{V}_y}{\partial \chi} f(\chi) - \frac{1}{2} \left\| \frac{\partial \mathcal{V}_y}{\partial \chi} \right\|_{\mathbf{Q}}^2 
    +\frac{1}{2} \| y(t) - h(\chi) \|_{\mathbf{R}^{-1}}^2 .
\end{equation}

Along the optimal trajectory, $\nabla \mathcal{V}_y(\hat{x}^*, t) = 0$ holds. Differentiating it yields the classical Mortensen observer:
\begin{equation}
    \label{eq:mortensen}
    \dot{\hat{x}}^* = f(\hat{x}^*) + \left( \nabla^2\mathcal{V}_y(\hat{x}^*, t) \right)^{-1} \frac{\partial h}{\partial x}(\hat{x}^*) ^\top\mathbf{R}^{-1} (y - h(\hat{x}^*)).
\end{equation}
Although computing $\nabla^2 \mathcal{V}_y$ requires solving the nonlinear HJB equation, which is generally intractable, this expression provides the \emph{local optimal behavior} that we aim to reproduce within the KKL framework.

\subsection{First-order Approximation of the MEE}

In the Mortensen filter framework, the observer state is composed of $z=(\hat x, \mathcal V_y)$, living in an infinite-dimensional space. The dynamics $\varphi$ correspond to the coupled system \eqref{eq:HJB}-\eqref{eq:mortensen}.
To interpret this within the structural framework of Proposition~\ref{prop:local_dyn_e}, consider the immersion mapping $T$ defined as
$$
T(x)=\begin{bmatrix}
    x \\ \mathcal W(x,\cdot)
\end{bmatrix},
$$
where $\mathcal W : \RR^{n_x} \times \RR^{n_x} \rightarrow \RR$ is the solution to the PDE
\begin{multline*}
    - \frac{\partial \mathcal W}{\partial x}(x,\chi)f(x)
    - \frac{\partial \mathcal W}{\partial \chi}(x,\chi)f(\chi) \\
    + \frac{1}{2} \| h(x) - h(\chi) \|_{\mathbf{R}^{-1}}^2 
    - \frac{1}{2} \left\| \frac{\partial \mathcal{W}}{\partial \chi}(x,\chi) \right\|_{\mathbf{Q}}^2
    =
    0,
\end{multline*}
subject to the boundary condition $\frac{\partial \mathcal W}{\partial \chi}(x,x)=0$.
In this setting, the reconstruction map $\tau$ is simply the projection onto the finite-dimensional state space $\RR^{n_x}$, \textit{i.e.}, $\tau(z) = \hat x$.

Consequently, the structural Jacobians satisfy:
$$
\frac{\partial \tau}{\partial z} = \begin{bmatrix} I_{n_x} & 0 \end{bmatrix}, \quad
\frac{\partial T}{\partial x}(x) = \begin{bmatrix} I_{n_x} \\ \frac{\partial \mathcal{W}}{\partial x} \end{bmatrix}.
$$
By applying formally the structural result of Proposition~\ref{prop:local_dyn_e} to this specific setting, we define the optimal effective gain $\Psi_{\text{opt}}$ and the corresponding optimal latent gain $\psi_{\text{opt}}$.
The general gain $\Psi(\hat x)$ in \eqref{eq:dyn_obs_exact_e} specializes to:
$$
    \Psi_{\text{opt}}(\hat x) = \psi_{\text{opt}}(\hat x) \frac{\partial T}{\partial x}(\hat x) = \psi_{\text{opt},1}(\hat x) + \psi_{\text{opt},2}(\hat x) \frac{\partial \mathcal{W}}{\partial x},
$$
where $\psi_{\text{opt}} = [\psi_{\text{opt},1}, \psi_{\text{opt},2}]$ denotes the partition of the latent gain corresponding to the state estimate component ($x$) and the value function component ($\mathcal{W}$), respectively.
Also, note that 
% $\frac{\partial^2 \tau}{\partial z^2} = 0$.
$\frac{\partial^2 \tau}{\partial z^2} = 0$ since $\tau$ is a linear projection.
Computing the Jacobian of $\varphi$ evaluated at 
% $z=T(x)$  and $y = h(\hat x)$ yields:
$z=T(\hat{x})$ and $y = h(\hat x)$ yields:
$$
    \left. \frac{\partial \varphi_1}{\partial z_1} \right|_{y=h(\hat x)} = \frac{\partial f}{\partial x}(\hat x) 
    % - P(\hat x)^{-1} C(\hat x),
    - P(\hat x)^{-1} C(\hat x)^\top \mathbf{R}^{-1} C(\hat x),
$$    
$$
    \left. \frac{\partial \varphi_1}{\partial z_2} \right|_{y=h(\hat x)} \!\!\! = \frac{\partial }{\partial z_2}\left\{
    \!\left( P(\hat x)\right)^{-1} \frac{\partial h}{\partial x}(\hat{x}) ^\top\mathbf{R}^{-1}\!\right\} \underbrace{(y - h(\hat x))}_{=0} = 0
$$
where $C(\hat x) = \frac{\partial h}{\partial x}(\hat x)$, and
$
P(\hat x) = \frac{\partial^2 \mathcal{W}}{\partial {\chi}^2}(\hat x, \hat x).
$
This gives:
$\psi_{\text{opt}}(\hat x) = \begin{bmatrix} \Psi_{\text{opt}}(\hat x) &  0 \end{bmatrix},$
where
\begin{equation*}%\label{eq:Psi_opt}
\Psi_{\text{opt}}(\hat x) = P(\hat x)^{-1} C(\hat x)^\top \mathbf{R}^{-1} C(\hat x).
\end{equation*}
Here, $P(x)$ corresponds to the Information Matrix (inverse of the estimation error covariance) and satisfies the following Partial Differential Riccati Equation (PDRE) on $\X$:
\begin{multline}
    \label{eq:PDRE}
    \mathcal{L}_f P(x) + P(x) A(x) + A(x)^\top P(x) \\+ P(x) \mathbf{Q} P(x) - C(x)^\top \mathbf{R}^{-1} C(x) = 0,
\end{multline}
where $\mathcal{L}_f P(x)$ denotes the Lie derivative of $P$ along $f$, and $A(x) = \frac{\partial f}{\partial x}(x)$.
Hence, we (almost) recover the observer suggested in \cite{sanfelice2015solution}.
\solutionswitch{For additional intuition, in \cite{galimberti2026}, we detail how the analogous observer structure is obtained in the linear case.}{For additional intuition, in Appendix~\ref{app:obs_linear_case}, we detail how the analogous observer structure is obtained in the linear setting.}

\subsection{Proposed Strategy}
\label{sec:strategy}

The analysis conducted in the previous sections reveals a complementarity between the KKL framework and the theory of optimal filtering.
On the one hand, the KKL observer ensures a global robust estimation property through the contraction of the latent dynamics $\varphi$ (Proposition \ref{prop:iss}). However, the selection of $\varphi$ is generally done heuristically, often leading to sub-optimal local performance.
On the other hand, the Mortensen filter defines the optimal estimation trajectory for a given cost functional \eqref{eq:cost_functional}. While its exact implementation is infinite-dimensional, its first-order approximation yields a tractable local behavior characterized by the gain matrix $K_{\text{opt}}(x) = P(x)^{-1} C(x)^\top \mathbf{R}^{-1}$, where $P(x)$ is the solution to the PDRE \eqref{eq:PDRE}.

Based on these observations, we propose a strategy to design a ``locally optimal'' KKL observer. The objective is to determine the parameters of the latent dynamics $\varphi$ and the associated transformation maps such that the local dynamics of the KKL observer match the optimal dynamics of the approximated Mortensen filter.

Specifically, recalling the local expansion in Proposition \ref{prop:local_dyn_e}, the KKL observer behaves locally as:
\begin{equation}\label{eq:kkl_local_dynamics_Psi}
    \dot{\hat x} \approx f(\hat x) + \Psi_{\text{KKL}}(\hat x) (x-\hat x(t)).
\end{equation}
Conversely, the optimal first-order filter evolves as:
\begin{equation}\label{eq:kkl_local_dynamics_psiopt}
    \dot{\hat x} \approx f(\hat x) +  {\Psi}_{\text{opt}}(\hat x) (x-\hat x(t)).
\end{equation}
Our goal is to identify the KKL components such that the effective correction term $\Psi_{\text{KKL}}$ mimics the optimal injection term $\Psi_{\text{opt}}(\hat x)$.
This turns the infinite-dimensional design problem into a parameter search problem. In the following section, we detail how Deep Learning is employed to solve this matching problem while satisfying the structural constraints of the KKL theory.

\section{Learning a contraction-based estimation} \label{sec:learning_method}

In this section, we shift our focus to the construction of practical realizations of KKL observers with local optimal steady-state performance.
We aim to learn mappings ${T}$, ${\tau}$, and latent dynamics ${\varphi}$ satisfying three core requirements:
\begin{itemize}
    \item[\texttt{R1})] 
    Adherence to the KKL framework, namely \eqref{eq:PDE}~and~\eqref{eq:tau_T_x};
    \item[\texttt{R2})] 
    Contraction in the latent space, satisfying \eqref{eq:contraction_latent_dynamics};
    \item[\texttt{R3})] 
    Local steady-state optimality, matching the effective gain $\Psi_{\text{KKL}}(\hat{x})$ in \eqref{eq:kkl_local_dynamics_Psi} with the optimal term $\Psi_\text{opt}(\hat{x})$ in \eqref{eq:kkl_local_dynamics_psiopt}.
\end{itemize}
Requirement \texttt{R3} relies on the information matrix $P(x)$ from the PDRE \eqref{eq:PDRE}. 
Since solving the PDRE is intractable for general nonlinear systems, we adopt a learning-based approach to approximate its inverse, $\hat{P}^{-1}$.

\subsection{Function Approximators}

We employ Multi-Layer Perceptrons (MLPs) to parameterize the immersion map $\hat{T}(x;\theta_T)$, the reconstruction map $\hat{\tau}(z;\theta_\tau)$, and the inverse information matrix $\hat{P}^{-1}(x;\theta_P)$, approximating $T$, $\tau$ and ${P}^{-1}$. The network parameters are collected in $\theta_T$, $\theta_\tau$ and $\theta_P$. 

For the latent dynamics $\hat{\varphi}(z,y;\vartheta)$, we seek for an architecture structurally enforcing the contraction property (\texttt{R2}) for any parameter $\vartheta$. 
Several stabilizing parametrizations have been proposed in both discrete and continuous time \citep{gu2022efficiently, revay2023recurrent, forgione2021dynonet, beik2024neural, zakwan2022robust}.

In this work, we adopt NodeRENs \citep{martinelli2023unconstrained}, which provide an unconstrained parametrization of continuous‑time contracting nonlinear systems, ensuring global convergence of the latent trajectories by design.

\subsection{Approximating the Inverse Information Matrix}

We train $\hat{P}^{-1}(x;\theta_P)$ prior to the estimator learning step. 
By left- and right-multiplying the PDRE \eqref{eq:PDRE} by $P^{-1}(x)$, we obtain an equivalent condition that depends only on $P^{-1}(x)$ itself.
Our goal is to find $\theta_P$ that minimize the Frobenius norm of the residual $\varrho$ of the transformed PDRE:
\begin{align}
    &\varrho(x, \theta_P) =  \notag \\
    &\quad -\mathcal{L}_f \hat{P}^{-1}(x; \theta_P) + \hat{P}^{-1}(x ; \theta_P) A^\top(x) + A(x) \hat{P}^{-1}(x ; \theta_P)  \notag \\
    &\quad - \hat{P}^{-1}(x ; \theta_P) C^\top(x) \mathbf{R}^{-1} C(x) \hat{P}^{-1}(x ; \theta_P) + \mathbf{Q}. 
    \label{eq:P_residual}
\end{align}
Given a dataset of state samples 
\begin{equation}
    \label{eq:dataset}
    \mathcal{D} = \{x_i \mid x_i\in\mathcal{X}\}_{i=1}^{N},
\end{equation}
the loss to minimize is
\begin{equation}
    \label{eq:optimization_P}
    \min_{\theta_P} 
    \;
    \frac{1}{N}
    \sum_{x \in\mathcal{D}}
    \big\| \varrho(x , \theta_P) \big\|_F^2 .
\end{equation}

\subsection{Learning the Locally Optimal Observer}

With $\hat{P}^{-1}$ fixed, we jointly train $(\theta_T, \theta_\tau, \vartheta)$ to satisfy \texttt{R1} and \texttt{R3}---the NodeREN architecture already guarantees \texttt{R2}. 
The overall loss evaluated over $\mathcal{D}$ is:
\begin{equation} 
    \label{eq:loss_L}
    \mathscr{L}(\theta_T, \theta_\tau,\vartheta) = \sum_{x\in \mathcal{D}} \big( \ell_{\text{pde}} + \ell_{\text{inv}} + \ell_{\text{opt}} \big),
\end{equation}
where the first term penalizes deviations from the PDE, while the other two promote the consistency between $\hat{T}$ and $\hat{\tau}$, and the optimal local behavior. Respectively, they are defined as:\footnote{
    With a slight abuse of notation, we use 
    $\hat{\Psi}_\text{KKL}(\hat{x}; \theta_T, \theta_\tau, \vartheta)$ 
    to denote the mapping $\Psi_\text{KKL}(\hat{x})$ in \eqref{eq:kkl_local_dynamics_Psi}
    but computed using the parametrized functions 
    $\hat{T}$, $\hat{\tau}$, and $\hat{\varphi}$ instead of 
    $T$, $\tau$, and $\varphi$.
}
\begin{align*}
    \ell_{\text{pde}} 
    &= \Big| \frac{\partial \hat{T}}{\partial x}(x;\theta_T) f(x) - \hat{\varphi}(\hat{T}(x;\theta_T), h(x) ; \vartheta)\Big|^2, \\
    \ell_{\text{inv}} 
    &= \left|\hat{\tau}(\hat{T}(x; \theta_T) ;\theta_\tau) - x \right|^2, \\
    \ell_{\text{opt}} 
    &= 
    \begin{aligned}[t]
        \Big| \hat{P}^{-1}(\hat{x};\theta_P) \frac{\partial h}{\partial x}^\top(\hat{x}) \, \mathbf{R}^{-1} \frac{\partial h}{\partial x}(\hat{x}) (x-\hat{x}) \\
        - \hat{\Psi}_{\text{KKL}}(\hat{x} ; \theta_T, \theta\tau, \vartheta )  (x - \hat{x}) \Big|^2.
    \end{aligned}
\end{align*}
For $\ell_{\text{opt}}$, the perturbed estimated state is $\hat{x} = \hat{\tau}(z_\text{pert}; \theta_\tau)$, where $z_\text{pert}$ is sampled from a ball $\mathcal{B}_r(\hat{T}(x;\theta_T))$. All parameters are optimized via stochastic gradient descent using automatic differentiation.

\section{Simulations}

This section illustrates the performance of the proposed learning-based contraction observer.
The goal of these academic examples is to demonstrate that 
(i) the mappings $\hat T$, $\hat \tau$, and $\hat \varphi$ can be jointly learned while satisfying the  requirements \texttt{R1}-\texttt{R3} of the KKL framework, and 
(ii) the resulting estimator exhibits the expected contraction and local steady-state optimality properties.

\subsection{Simulation Setup}
We evaluate the proposed learning-based observer on two benchmark nonlinear systems, both of which are two-dimensional, oscillatory, and nonlinear.

The first considered system is the Van der Pol oscillator
\begin{align*}
\Sigma_\text{VdP} \,
\left\{ \;
\begin{aligned}
    \dot{x}_1(t) &= x_2(t) + w_1(t),\\
    \dot{x}_2(t) &= \left(1- x_1^2(t)\right) \, x_2(t) -  x_1(t) + w_2(t),\\
    y(t) &= x_1(t) + v(t).
\end{aligned}
\right.
\end{align*}
The second system is the reverse Duffing oscillator described by
\begin{align*}
\Sigma_\text{duff}\,
\left\{ \;
\begin{aligned} %[t]
    \dot{x}_1(t) &= x_2^3(t) + w_1(t), \\
    \dot{x}_2(t) &= -x_1(t) + w_2(t),  
\end{aligned}    
\right.\qquad 
y(t) = x_1(t) + v(t) .
\end{align*}

\subsection{Training Domain and Datasets}
We consider a compact set $\mathcal{X}$ that will contain all the trajectories considered for evaluation. For $\Sigma_{\text{VdP}}$, we set $\mathcal{X}_\text{VdP} = [-2.5, 2.5] \times [-3.5, 3.5]$, and for $\Sigma_{\text{duff}}$, we consider $\mathcal{X}_\text{duff} = [-4, 4]^2$.

For both benchmarks, the training dataset \eqref{eq:dataset} contains $N = 5\times 10^4$ state points.
Over this dataset, we first train $\hat{P}^{-1}(x;\theta_P)$ via the minimization problem \eqref{eq:optimization_P} where we select matrices $\mathbf{Q} = I_2$ and $\mathbf{R} \in \{1, \, 10^{-2} \}$. 
Then, we train our proposed observer architectures, for which the loss \eqref{eq:loss_L} is evaluated at each iteration.
All networks are trained using stochastic gradient descent with the Adam optimizer.  
While evaluating the local optimality term of the loss, i.e., $\ell_\text{opt}$, perturbed latent states $z_\text{pert}$ are sampled from a ball $\mathcal{B}_r(T(x))$ of radius $r= 10^{-1}$.

\subsection{Results}
All simulations, results, as well as all the chosen hyperparameters can be found at 
\url{https://github.com/ClaraGalimberti/locally_optimal_KKL_observers}.

\begin{figure}[!tb]
	\centering
	\includegraphics[width=0.9\linewidth]{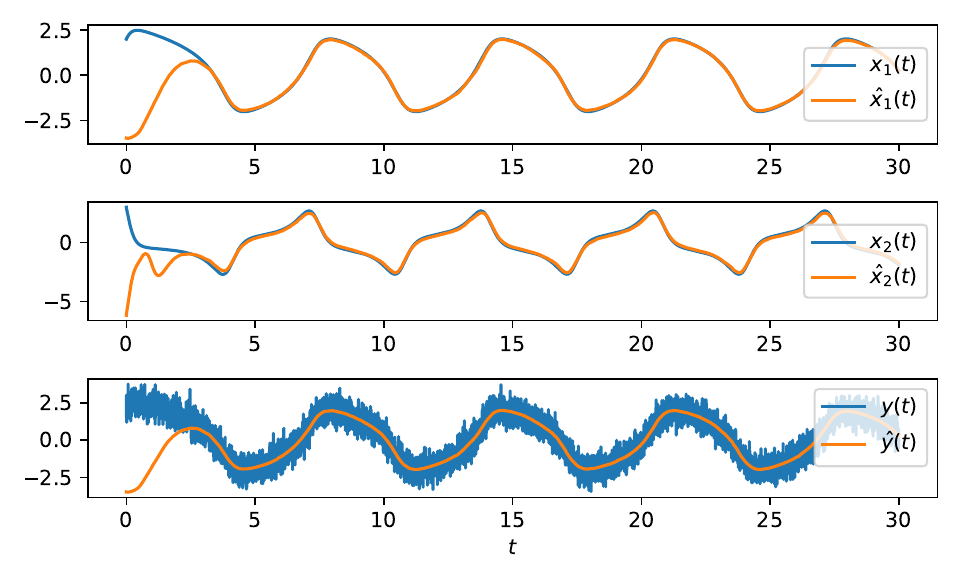}
	\caption{True states $x(t)$, estimated states $\hat{x}(t)$, and noisy measurements $y(t)$ for the Van der Pol system.}
	\label{fig:states_vdp}
\end{figure}

\begin{figure}[!tb]
	\centering
\includegraphics[width=0.9\linewidth]{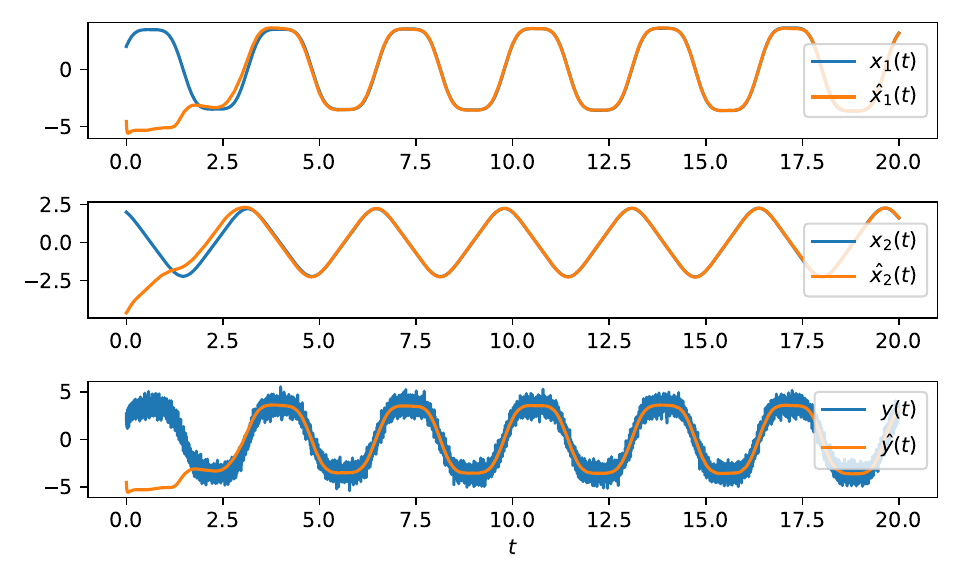}
	\caption{True states $x(t)$, estimated states $\hat{x}(t)$, and noisy measurements $y(t)$ for the inverse Duffing oscillator.}
	\label{fig:states_duff}
    %\mn{c'est quand même mieux ;-)} \clara{:)}
\end{figure}

We evaluate the trained observers with $\mathbf{R}=1$ on a randomly chosen trajectory within the domain $\mathcal{X}$, under measurement noise, i.e., $w(t)=0$ and $v(t) \sim \mathcal{N}(0,0.5^2)$.
Figures~\ref{fig:states_vdp} and~\ref{fig:states_duff} show the true states $x(t)$, the state estimates $\hat{x}(t)$, and the noisy measurements $y(t)$ for $\Sigma_\text{VdP}$ and $\Sigma_\text{duff}$, respectively.

We observe that predicted states maintain a stabilizing behavior even when initialized far from the true initial conditions, thanks to the contraction property of $\hat{\varphi}$. Furthermore, we observe that the learned estimators accurately track the states despite significant measurement noise.

\begin{figure}[!tb]
	\centering
	\begin{minipage}{0.5\linewidth}
		\includegraphics[width=\linewidth]{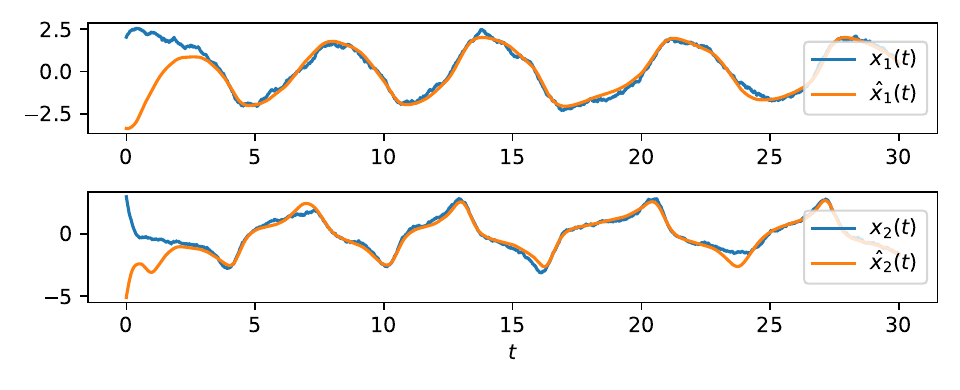}
	\end{minipage}%
	\begin{minipage}{0.5\linewidth}
		\includegraphics[width=\linewidth]{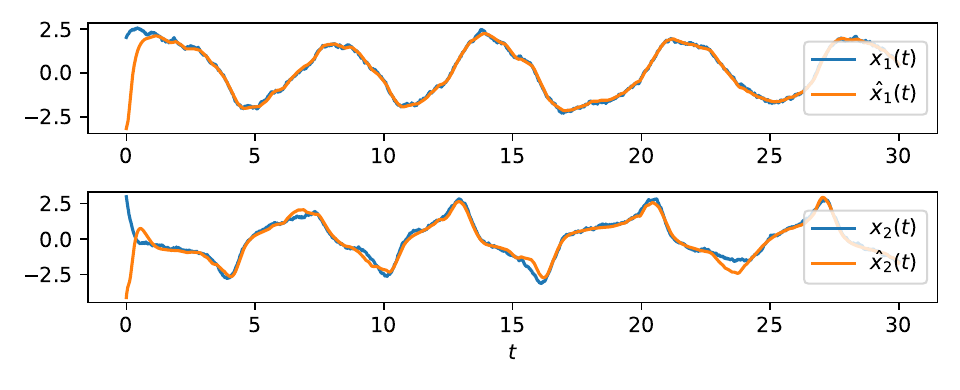}
	\end{minipage}%
    \caption{Comparison of learned observers for $\Sigma_\text{VdP}$ trained with $\mathbf{Q}=I$ and different values of $\mathbf{R}$. Trajectories were generated with $v(t)=0$ and $w(t) \sim \mathcal{N}(0,0.25^2)$. \textit{Left:} $\mathbf{R}=1$. \textit{Right:} $\mathbf{R}=10^{-2}$. % TODO \johan{design: $ \mathbf{Q}$ pour $w$, $\mathbf{R}$ pour $v$}
    }
	\label{fig:vdp_different_weights}
\end{figure}

\begin{figure}[!tb]
	\centering
	\begin{minipage}{0.5\linewidth}
		\includegraphics[width=\linewidth]{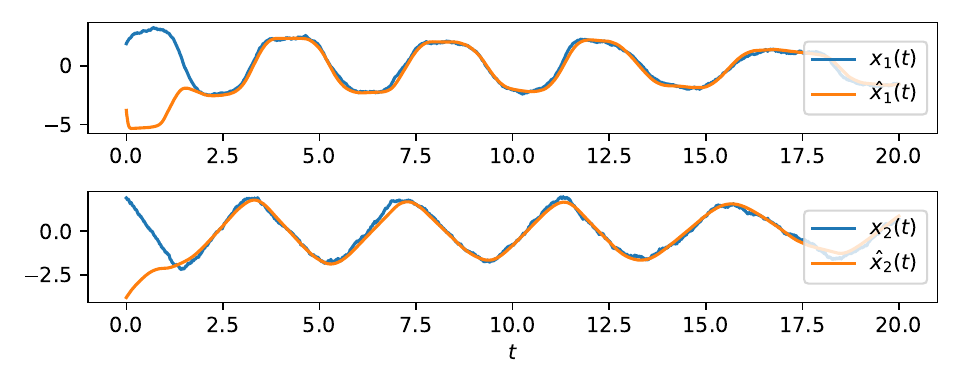}
	\end{minipage}%
	\begin{minipage}{0.5\linewidth}
		\includegraphics[width=\linewidth]{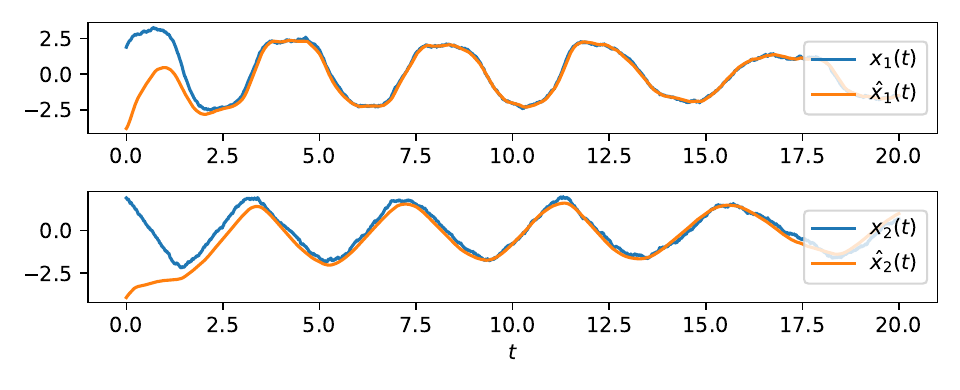}
	\end{minipage}%
    \caption{Comparison of learned observers for $\Sigma_\text{duff}$ trained with $\mathbf{Q}=I$ and different values of $\mathbf{R}$. Trajectories were generated with $v(t)=0$ and $w(t) \sim \mathcal{N}(0,0.25^2)$. \textit{Left:} $\mathbf{R}=1$. \textit{Right:} $\mathbf{R}=10^{-2}$.}
	\label{fig:duff_different_weights}
\end{figure}

We next evaluate our observers trained with different values of $\mathbf{R}$, under process noise only. Figures~\ref{fig:vdp_different_weights} and~\ref{fig:duff_different_weights} show the evolution of $x(t)$, $\hat{x}(t)$, and  measurements $y(t)$. In both figures, the left panels correspond to the case $\mathbf{R}=1$, while the right panels show results for $\mathbf{R}=0.01$.

In all simulations, the noise profile does not change: $v(t)=0$ and $w(t)\sim \mathcal{N}(0,0.25^2)$. We find that training with a smaller $\mathbf{R}$ leads the observer to track the noisy state dynamics more closely, while a larger $\mathbf{R}$ makes the observer rely more on the available model.

\section{Conclusion}

We proposed a way to learn a KKL observer that is globally contractive and locally mimics the behavior of the 
minimum energy (Mortensen) estimator. Our training objective combines the KKL PDE, and a local optimality condition, while the use of NodeRENs guarantees contraction by design. Simulations on two nonlinear systems demonstrate the good performance on academic examples where the learned observers behave reliably and handle noise well. Next steps include 
% \emph{making the connection with the Mortensen observer more explicit[¿?]},
establishing approximation bounds for the learned models, and testing the approach on systems where standard observers may perform poorly. Another route of investigation is the derivation of an equivalent 
theory for discrete-time dynamics. 

\bibliography{ifacconf}

\clearpage
\appendix

\section{Review of the linear case}
\label{app:obs_linear_case}
It is instructive to instantiate the proposed framework for a linear system $\dot x = Ax, \ y = Cx$.
Consistent with Section \ref{sec:optimal_behavior}, let $P \succ 0$ be the steady-state information matrix, solution to the Algebraic Riccati Equation:
\begin{equation*}
     PA + A^\top P +P\mathbf{Q}P  - C^\top \mathbf{R}^{-1} C = 0.
\end{equation*}
The optimal observer is $\dot{\hat x} = A \hat x + K_{\text{opt}}(y-C\hat x)$, with $K_{\text{opt}} = P^{-1}C^\top \mathbf{R}^{-1}$.
On the other hand, a linear KKL observer $\dot z = Fz + Gy$, $\hat x= T^{-1}z$ relies on the Sylvester equation $TA = FT + GC$.
Substituting $FT = TA - GC$ into the dynamics of $\hat x = T^{-1}z$, we obtain:
\begin{align*}
    \dot{\hat x} &= T^{-1}(FT \hat x + G y) \nonumber = T^{-1}(TA - GC)\hat x + T^{-1}G y \nonumber \\
    &= A\hat x + \underbrace{T^{-1}G}_{\Psi} (y - C\hat x).
\end{align*}
Comparing the two structures, the KKL observer matches the optimal behavior if and only if the effective gain satisfies $\Psi = K_{\text{opt}}$.
Substituting $G = T K_{\text{opt}}$ back into the Sylvester equation yields the similarity condition:
\begin{equation*}
    F = T(A - K_{\text{opt}}C)T^{-1}.
\end{equation*}
This implies that for linear systems, the latent dynamics $F$ cannot be chosen arbitrarily; they must share the spectrum of the optimal closed-loop error dynamics.

\section{Proof of Proposition~\ref{prop:iss} }

    Let $z^*(t) = T(x(t))$. Differentiating $z^*$ along the trajectories of the disturbed system \eqref{eq:sys} yields:
    \begin{align*}
        \dot{z}^* &= \frac{\partial T}{\partial x}(x) (f(x) + w) \\
        &= \varphi(T(x), h(x)) + \frac{\partial T}{\partial x}(x)w \\
        &= \varphi(z^*, y - v) + \frac{\partial T}{\partial x}(x)w.
    \end{align*}
    The observer dynamics are described by \eqref{eq:phi_dynamics}. %$\dot{\hat{z}} = \varphi(\hat{z}, y)$, \clara{with initial condition $z_0$}. 
    The error %$\tilde{z} = \hat{z} - z^*$ 
    $\tilde{z} = z - z^*$ 
    evolves as:
    $$ 
    \begin{aligned}
          \dot{\tilde{z}} &= \varphi(z, y) - \varphi(z^*, y) + \Delta(t) 
          \\
        \Delta(t) &= \varphi(z^*, y) - \varphi(z^*, y-v) - \tfrac{\partial T}{\partial x}(x) w . 
    \end{aligned}
    $$
    Using the contraction of $\varphi$ (Assumption~\ref{ass:contraction}) and the Lipschitz properties of $\varphi$ and $T$, standard ISS arguments yield the bound on $\tilde{z}$. The bound on $\tilde{x}$ follows from $|\hat{x}-x| = |\tau(z) - \tau(z^*)| \le L_\tau |\tilde{z}|$.

\section{Proof of Proposition~\ref{prop:local_dyn_e}}
Let $z^* = T(x)$. We proceed by expanding the observer dynamics 
$\dot{\hat{x}} = \frac{\partial \tau}{\partial z}(z^*+e) \varphi(z^*+e, h(x))$
using Taylor's expansion.
First, since $\varphi$ is $C^2$ and $\tau$ is $C^3$, there exist remainder functions $\rho_\varphi(e,x)$ and $\rho_\tau(e,x)$ such that:
\begin{align*}
    \varphi(z^*+e, h(x)) &= \varphi(z^*, h(x)) + \frac{\partial \varphi}{\partial z}(z^*, h(x)) e + \rho_\varphi(e,x), \\
    \frac{\partial \tau}{\partial z}(z^*+e) &= \frac{\partial \tau}{\partial z}(z^*) + \frac{\partial^2 \tau}{\partial z^2}(z^*) e + \rho_\tau(e,x),
\end{align*}
satisfying $|\rho_\varphi(e,x)| \le \frac{L_{\varphi,2}}{2}|e|^2$ and $|\rho_\tau(e,x)| \le \frac{L_{\tau,3}}{2}|e|^2$ for all $(e,x)$ in $\RR^{n_z}\times\X$.
Using the PDE \eqref{eq:PDE} along with the fact that $\frac{\partial \tau}{\partial z}(z^*)\frac{\partial T}{\partial x}(x)=I$ (obtained by differentiating \eqref{eq:tau_T_x}), we obtain:
\begin{equation}\label{eq:proof_step1}
    \dot{\hat{x}} = f(x) + M(x) e + \rho_1(x,e),
\end{equation}
where 
$M(x) = \frac{\partial \tau}{\partial z}(z^*) \frac{\partial \varphi}{\partial z}(z^*, h(x)) + \frac{\partial^2 \tau}{\partial z^2}(z^*) \varphi(z^*, h(x))$,
and $\rho_1$ contains all higher-order and cross terms. Using the bounds on derivatives, $|\rho_1(x,e)| \le k_1 |e|^2$ for some $k_1>0$.

Moreover, note that $\hat{x} = \tau(z^*+e) = x + \frac{\partial \tau}{\partial z}(z^*)e + \rho_x(e,x)$ with $|\rho_x| \le \frac{L_{\tau,2}}{2}|e|^2$.
Applying a Taylor expansion to $f$ yields:
\begin{equation*}
    f(\hat{x}) = f(x) + \frac{\partial f}{\partial x}(x) \left( \frac{\partial \tau}{\partial z}(z^*)e + \rho_x(e,x) \right) + \rho_f(\hat{x}-x),
\end{equation*}
where $|\rho_f(\delta)| \le \frac{L_{f,2}}{2} |\delta|^2$. This can be rewritten as:
\begin{equation}\label{eq:proof_step2b}
    f(x) = f(\hat{x}) - \frac{\partial f}{\partial x}(x) \frac{\partial \tau}{\partial z}(z^*)e + \rho_2(x,e),
\end{equation}
where the combined remainder satisfies $|\rho_2(x,e)| \le k_2 |e|^2$, for some $k_2>0$.
Also, replacing evaluations at $x$ with evaluations at $\hat{x}$ in the linear terms of \eqref{eq:proof_step1} and \eqref{eq:proof_step2b}:
\begin{equation}\label{eq:proof_step3}
    \left( M(x) - \frac{\partial f}{\partial x}(x) \frac{\partial \tau}{\partial z}(z^*) \right) e = \psi(\hat{x}) e + \rho_3(x,e),
\end{equation}
where $|\rho_3| \le L_{\psi} |\hat{x}-x| |e| \le k_3 |e|^2$ (using $|\hat{x}-x| \le L_\tau |e|$).

Substituting \eqref{eq:proof_step2b} into \eqref{eq:proof_step1} and using \eqref{eq:proof_step3}, we get:
\begin{equation*}
    \dot{\hat{x}} = f(\hat{x}) + \psi(\hat{x}) e + \mathcal{R}_1(x,e),
\end{equation*}
where $\mathcal{R}_1 = \rho_1 + \rho_2 + \rho_3$ satisfies $|\mathcal{R}_1| \le (k_1+k_2+k_3)|e|^2$.

To express this in terms of the estimation error $\tilde{x} = \hat{x} - x$, we decompose the latent error $e$ as follows:
\[
e = (T(\hat{x}) - T(x)) + (z - T(\hat{x})).
\]

Using the first-order Taylor expansion of $T$ around $\hat{x}$:
\[
T(\hat{x}) - T(x) = \frac{\partial T}{\partial x}(\hat{x}) (\hat{x} - x) + \rho_T(x,e),
\]
where $|\rho_T(x,e)| \le \frac{L_{T,2}}{2} |\hat{x}-x|^2 \le k_T |e|^2$ (using $|\hat{x}-x| \le L_\tau |e|$).
Injecting this approximation into the dynamics yields:
\begin{equation*}
    \dot{\hat{x}} = f(\hat{x}) + \psi(\hat{x}) \!\left[ \tfrac{\partial T}{\partial x}(\hat{x}) (\hat{x}-x)
    + z - T(\hat{x})\right] + \mathcal{R}_1 + \psi(\hat{x})\rho_T.
\end{equation*}
Identifying $\Psi(\hat{x}) = -\psi(\hat{x}) \frac{\partial T}{\partial x}(\hat{x})$, we obtain \eqref{eq:dyn_obs_exact_e} where the  remainder term is:
\[
\mathcal{R}(x, e) = \mathcal{R}_1(x, e) + \psi(\hat{x})\rho_T(x,e) + \psi(\hat x)(z-T(\hat x)).
\]

We now bound $\mathcal R(x,e)$. From the previous estimates,
\[
|\mathcal R_1(x,e)| \le k_1' |e|^2, 
\qquad
|\psi(\hat x)\rho_T(x,e)| \le k_2' |e|^2,
\]
for some $k_1',k_2'>0$ depending on Assumption~\ref{ass:bounds}.
Next, decompose
$\psi(\hat x) = \psi(x) + \big(\psi(\hat x) - \psi(x)\big)$, so that
\[
\psi(\hat x)(z-T(\hat x))
= \psi(x)(z-T(\hat x)) + \big(\psi(\hat x) - \psi(x)\big)(z-T(\hat x)).
\]

Since $x\in\X$ and $\X$ is compact, continuity of $\psi$ implies that there exists a constant $\bar C_\psi>0$ such that
$|\psi(x)(z-T(\hat x))| \le \bar C_\psi |z-T(\hat x)|$.
Moreover, by Lipschitz continuity of $\psi$ and the bound $|\hat x - x|\le L_\tau |e|$, we have
$|\psi(\hat x) - \psi(x)| \le L_\psi |\hat x - x| \le L_\psi L_\tau |e|$.
Thus,
\begin{align*}
    \big|(\psi(\hat x) - \psi(x))(z-T(\hat x))\big|
    &\le L_\psi L_\tau |e|\,|z-T(\hat x)| \\
    &\le \frac{L_\psi L_\tau}{2}\big(|e|^2 + |z-T(\hat x)|^2\big),
\end{align*}
where we used Young's inequality.
Collecting all bounds, we obtain
\[
|\mathcal R(x,e)|
\le c_1 |e|^2 + c_2 |z-T(\hat x)| + c_3 |z-T(\hat x)|^2,
\]
for suitable positive constants $c_1,c_2,c_3$ depending on the bounds in Assumption~\ref{ass:bounds}. 
This proves \eqref{eq_BoundRemainder} and concludes the proof.

\end{document}